\documentclass[prl,twocolumn]{revtex4}
\usepackage[dvips]{graphicx}          
\usepackage{amsmath}

\begin{document}
\title{Kauffman networks with threshold functions}
\author{Florian Greil and Barbara Drossel} 
\affiliation{Institut f{\"u}r Festk{\"o}rperphysik,  TU Darmstadt,
Hochschulstra{\ss}e 6, 64289 Darmstadt, Germany }
\date{\today}
\pacs{89.75.Hc, 05.70.Fh, 05.65.+b}
\begin{abstract}
We investigate Threshold Random Boolean Networks with $K = 2$
inputs per node, which are equivalent to Kauffman networks, with only
part of the canalyzing functions as update functions.
According to the simplest consideration these networks should be
critical but it turns out that they show a rich variety of behaviors,
including periodic and chaotic oscillations.
The results are supported by analytical calculations and computer
simulations.
\end{abstract} 

\newcommand{\up}{\uparrow}
\newcommand{\down}{\downarrow}

\maketitle

Random Boolean networks (RBN) were introduced by S.~Kauffman in 1969
\cite{kauffman:metabolic, kauffman:homeostasis} to model the dynamics of
genetic and metabolic networks \cite{kauffman:random}, but they are also
used in a social and economic context \cite{alexander:random,
paczuski:self-organized}, and for neural networks. Although Boolean
models represent a strong simplification of the far more complex
reality, there exist several examples where the modelling of a genetic
network by Boolean variables captures correctly the essential dynamics
of the system \cite{bornholdt:less,li:yeast,albert:topology}. For this
reason, the study of RBNs remains an important step on the way towards
understanding real networks.

A random Boolean network is a directed graph with randomly chosen
links between $N$ binary nodes. We denote the state of a node with
$\sigma_i = \pm 1$ (we call $+1$ ``on'' and $-1$ ``off''), and the
number of inputs per node with $K$. Each node $i$ is assigned at random
an update function $f_i$. In this paper, we focus on the case $K=2$ and
on threshold functions 
\begin{equation}
f_i = {\rm sign} \left( \sum_{j} \sigma_j c_{ij} \right) \equiv {\rm
  sign} \left( s_i \right) \,
 \label{TrbnDynamics}
\end{equation}
where the sum is taken over the two input nodes for node $i$, and
$c_{ij}=-1$ for \emph{inhibitory} connections and $c_{ij}=1$ for
\emph{excitatory} connections. (This version of RBN was used for
instance in \cite{rohlf:criticality}.) A connection is excitatory with
probability $p_+$ and inhibitory with probability $1-p_+$, leading to
the following table (Tab.~\ref{TruthK2}).
\begin{table}[htb]
\begin{tabular*}{0.45\textwidth}{lp{1.5cm}p{1.5cm}p{1.5cm}p{1.5cm}}
\hline 
Input             &  $f_{7}$ & $f_{11}$ & $f_{13}$ &$f_{14}$ \\ 
\hline
$(\down,\down)$ & $\up$   &$\up$   &$\up$   &$\down$ \\
$(\down,\up)$   & $\up$   &$\up$   &$\down$ &$\up$   \\
$(\up,\down)$   & $\up$   &$\down$ &$\up$   &$\up$   \\
$(\up,\up)$     & $\down$ &$\up$   &$\up$   &$\up$   \\
\hline 
probability &  $(1-p_+)^2$ & $p_+(1-p_+)$ & $p_+(1-p_+)$ & $p_+^2$ \\
\hline
\end{tabular*}
\caption{The four possible update functions for the model used in this
paper.  The input configuration is given in the first column, with $\up$
denoting $\sigma_i = 1$ and $\down$ denoting $\sigma_i = -1$.  The last
row gives the probability for each function, the top row gives the name
of the function according to the Kauffman model. 
\label{TruthK2}}
\end{table}

In agreement with other authors, we define sign$(0) = 1$. Threshold
functions are used not only in the context of neural networks, but also
in models for genetic networks \cite{li:yeast, rohlf:criticality,
bornholdt:robustness}.  These functions represent four of the 12
canalyzing update functions of Kauffman networks. Canalyzing functions
are those non-frozen functions where at least one value of a given input
can fix the output of a node, irrespective of the value of the second
input.  All nodes are updated in parallel according to the rule
\begin{equation}
\sigma_i(t+1) = f_i(\{\sigma_j(t)\}) 
\equiv f_i(\sigma_{i_1}(t),\sigma_{i_2}(t))\, .
\end{equation}
Node $i$ depends on the nodes $j$, namely on node $i_1$ and $i_2$.

The configuration of the system~$\vec{\sigma}
\equiv \{\sigma_1, \ldots, \sigma_N \}$ performs a trajectory in
configuration space. As the state space is finite and the dynamics is
discrete, some states will occur again. If a \emph{cycle}
in state space has a set of transient states leading to it, 
it is called an \emph{attractor}.

Kauffman classified the dynamics of RBNs according to whether it is
\emph{chaotic} or \emph{frozen} or \emph{critical} (as described in
the review \cite{aldana-gonzalez:boolean}). In a chaotic network, a
perturbation at one node propagates on an average to more than one
node, leading to long attractors. In a frozen network, such a
perturbation propagates on an average to less than one node, and in
the thermodynamic limit $N\to \infty$ only a finite number of nodes
are not constant after a certain transient time. Critical networks are
at the boundary between these two types of behavior, with a
perturbation of one node propagating on an average to one other
node. Therefore the difference between two almost identical initial
states increases like a power law in time. The number of nodes that
are not frozen on all attractors increases in a critical network as a
power law $\sim N^{2/3}$ of the system size $N$, as was found
numerically in \cite{socolar:scaling} and analytically in
\cite{samuelsson:superpolynomial,kaufman:scaling}.

It is the aim of this paper to show that this classification
breaks down for the simple class of RBNs considered here. We find a
much richer dynamical behavior with not only a frozen and a chaotic phase,
but also with two types of oscillating phases and several critical
points. 

Let us first apply the criticality condition in its simplest version:
For all four update functions, the probability that the output changes
if one input spin is flipped, is 1/2. Since each node is on an average
the input to two other nodes, a perturbation at one node propagates on
an average to one other node, and we should expect the model to be
critical. This is in agreement with the finding that $K\!=\!2$-RBNs that
contain only canalyzing update functions (but all of them with the
same probability) are critical \cite{paul:properties}.  However, this
simple argument is based on the assumption that all four possible
input configurations occur equally often, which may be true at the
beginning of a simulation run, but may be wrong already after one
timestep.  For this reason, Moreira and Amaral \cite{moreira:canalizing}
argued that the calculation should be performed such that the input
configurations are weighted with their frequencies in the stationary
state. 

Let us therefore next apply the rule given by Moreira and Amaral
and let us determine for what values of $p_+$ it predicts that the model is
frozen, critical, or chaotic. We will then see later that the result
is still not correct for all parameter values.

\begin{figure}[htb!]
\includegraphics[angle=-90,width=0.45\textwidth]{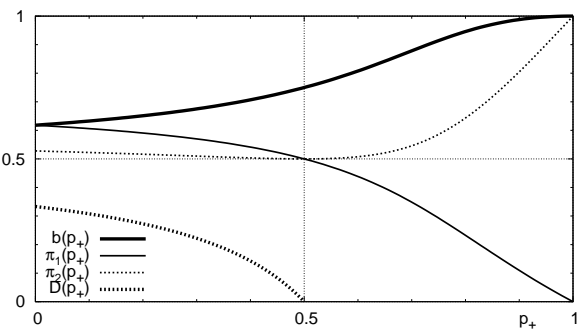}
\caption{The functions $b(p_+), \pi_1(p_+), \pi_2(p_+)$ and
the stationary value $D(p_+)$ vs{.} $p_+$.} \label{SketchBPi} 
\end{figure}

We denote with $b_t$ the proportion of nodes in state $\sigma_i= +1$
at time $t$. In the thermodynamic limit, it changes deterministically
according to
\begin{eqnarray}
b_{t+1} &=& 1 - \big[ b_t^2 \left(1-p_+\right)^2 +
  \left(1-b_t\right)^2 p_+^2 + \nonumber \\
 &&  2 b_t \left(1-b_t\right) p_+ \left(1-p_+\right)\big] \, .
 \label{TimeEvolutionB}
\end{eqnarray}
The expression in the square brackets is the probability that an input
combination leads to $s_i = -2$. 
In the stationary state, we have $b_{t+1} = b_t = b$ with 
\begin{equation}
b(p_+) = \frac{4p_+^2-2p_+ -1\! \pm \sqrt{5 - 12 p_+\! + 8
p_+^2}}{2(1-2p_+)^2}\, . \label{SolBranch}
\end{equation}
The sign in the numerator has to be chosen such that $b \in [0,1]$,
therefore only the positive branch remains, see
Fig.~\ref{SketchBPi}. For $p_+=1/2$, the denominator vanishes, and the
stationary solution of Eq.~(\ref{TimeEvolutionB}) is $b=3/4$. For $p_+ =
1$, we have $b=0$ and $b=1$, with the first solution being obviously
unstable as it is destroyed by one node in the state
$\sigma_i=+1$. The second solution is a stable fixed point of the
dynamics.  For $p_+=0$, we have $b=(-1+\sqrt{5})/2$.  

The mean number of nodes to which a perturbation at one node propagates
is in the stationary state given by $2\pi_1$, with $\pi_1$ being the
probability that a node changes its state when one input is flipped.  We
obtain it by adding up the probabilities for those input configurations
which allow a transition between an output $+1$ and $-1$ and vice versa.
This is true for half of the input configurations leading to $s_i = 0$
(the first 4 terms in the following equation) and for all input
configurations for which $s_i=-2$ (the last 4 terms):
\begin{eqnarray}
\pi_1 &=& (1-p_+)(1-b)(1-p_+)b  + p_+ b p_+ (1-b)    +\nonumber \\
      && p_+ b (1-p_+)b       + (1-p_+)(1-b)p_+(1-b) +\nonumber \\
      && (1-p_+) b (1-p_+) b  + p_+ b (1-p_+)(1-b)   +\nonumber \\
      && (1-p_+)(1-b) p_+ b   + p_+(1-b)p_+(1-b)      \nonumber \\
\Rightarrow \pi_1 &=& b + p_+ - 2 b p_+ \label{Pi1}
\end{eqnarray}
For $p_+=1/2$, we obtain $\pi_1 = 1/2$, for $p_+=1$, we obtain $\pi_1 =
0$. For $p_+=0$, we obtain $\pi_1=b \simeq 0.618$. We therefore conclude
that the model is in the frozen phase for $p_+ > 1/2$, that it is
critical for $p_+=1/2$, and chaotic for $p_+<1/2$. 

The same result is obtained by calculating the stationary value of the
Hamming distance between two identical network realizations. 
The  Hamming distance is the fraction of nodes for two configurations
$\vec \sigma, \tilde {\vec \sigma}$ that have different values:
$D=(4N)^{-1} \sum_{i=1}^{N} (\sigma_i - \tilde \sigma_i)^2$.  If we
denote with $\pi_2$ the probability that a node changes its state when
both inputs are flipped, the time evolution of $D$ is given by
\begin{equation}
D_{t+1} = 2 D_t (1-D_t) \pi_1 + D_t^2 \pi_2 \, .
\label{HammingK2}
\end{equation}
$\pi_2$ in the stationary state is obtained by summing all 8 combinations
leading to $s_i \in \{\pm 2\}$. It can be written as
\begin{equation}
\pi_2 = 1 - 2 b (1\!-\!2p_+)^2\!+ 2 b^2 (1\!-\!2p_+)^2\!- 2p_+ + 2 p_+^2
\label{Pi2}
\end{equation}
and is 1/2 for $p_+ = 1/2$ and 1 for $p_+=1$. If $D_t$ is very small, we
have 
\begin{equation}
D_{t+1} \simeq 2 D_t \pi_1 \, ,
\label{HammingK2_small}
\end{equation}
which allows for the growth of a small perturbation if $2\pi_1>1$ or
$p_+<1/2$, in agreement with our result above. The transition from a
stationary value $D=0$ to a stationary value $D>0$ occurs at the same
point. 

\begin{figure}[htb!]
\includegraphics[width=0.475\textwidth]{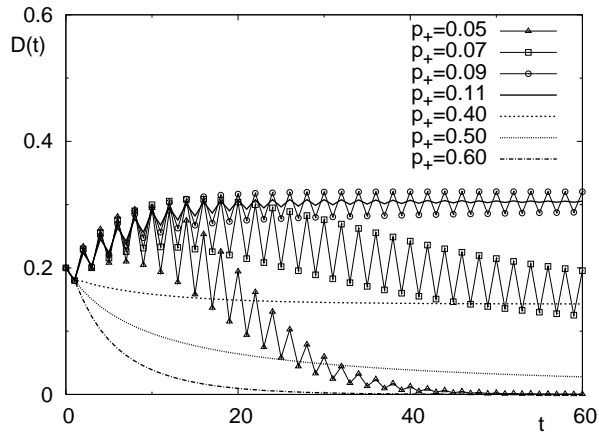}
\caption{The time evolution of the Hamming distance $D$ for different
values of $p_+$ when $b$ is not stationary. The curves are calculated
according to Eqs.~(\ref{TimeEvolutionB}), (\ref{HammingK2}) starting from
$D_0=0.2$ and $b_0=0.5$.  \label{TimeEvolveD} }
\end{figure}

Fig.~\ref{TimeEvolveD} shows $D_t$ for different values of $p_+$ and for
given initial conditions. One can see that $D_t$ approaches 0 for large
times if $p_+>0.5$. Furthermore, one can see that $D_t$ oscillates with
period 2 for the smaller values of $p_+$. This oscillation is an
indication that the dynamics in the ``chaotic'' phase has some
structure, which shall be investigated in the following.

\begin{figure}[bht!]
\includegraphics[width=0.3\textwidth]{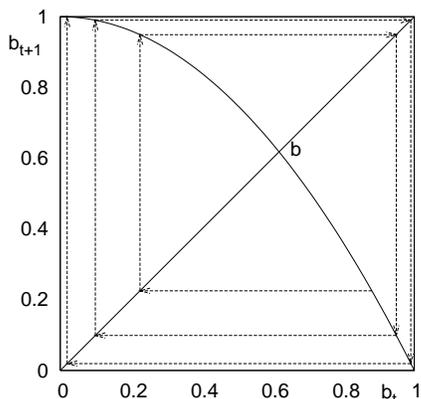}
\caption{The map $b_t$ vs. $b_{t+1}=1-b_t^2$ for $p_+=0$. The fixed
point~$b$ is unstable as depicted by a sample trajectory.}
\label{SketchMapping} 
\end{figure}

Let us therefore have a closer look at the supposedly chaotic phase
$p_+<1/2$.  We will see that the dynamics is not chaotic at all for
sufficiently small $p_+$. The considerations that have lead to our
simple phase diagram are flawed. The reason is that we have assumed that
$b$ becomes for large times stationary for all $p_+$.  In order to see
that this need not be the case, let us first look at the situation where
$p_+=0$: We then have $b_{t+1} = 1- b_t^2$. This is a one-dimensional
map shown in
Fig.~\ref{SketchMapping}.
The fixed point is unstable! Instead of having a stationary point with a
constant proportion of nodes in the two states, the system oscillates
between a configuration where all nodes are switched on and a state
where all nodes are switched off. This is not chaotic dynamics at all,
but very stable dynamics. 
In order to determine the range of $p_+$values, for which the fixed
point value of $b$ is unstable, we performed a linear stability analysis. 
The ansatz $b_{t+1}(b + \delta b_t) = b + \delta b_{t+1}$ leads in
linear order in $\delta b$ and for $p_+<1/2$ to
\begin{align}
\delta b_{t+1}&= -2 \left( b_t \left(1-2p_+\right)^2 
 + \left(1-2p_+\right) p_+ \right) \delta b_t \nonumber \\
&= \left( 1- \sqrt{5-12 p_+ + 8 p_+^2} \right) \delta b_t 
 =: M \cdot \delta b_t
\end{align}
In the last step we used Eq.~(\ref{SolBranch}).  The fixed point is stable
if the real part of $M$ is smaller than 1, which is the case if
\begin{equation}
p_+ > (3-\sqrt{7})/4 \equiv
p_{cb} \approx 0.0886 \, .\label{instability}
\end{equation}
Only above this value does the system have a stationary state with
constant proportions of nodes being ``on'' and ``off''. 

We finally investigate in more detail the region $p_+ < p_{cb}$, where
the proportion of ``on'' and ``off'' nodes oscillates with period
2. For $p_+=0$, every node oscillates with period 2, and we have a
global attractor of period 2. This need not necessarily be the case if
$b$ oscillates with period 2. The attractor could be much larger,
while the proportion of off and on nodes oscillates still with period
two.  In order to determine for which parameters an attractor with
period 2 is stable, we performed again a linear stability analysis,
but now for two time steps together.  We assume that the system is on
an attractor of length 2.  Let there be every even time step a
proportion~$x$ of ``on''-nodes and every odd step a proportion
$y$. The time evolution is given by Eq.~(\ref{TimeEvolutionB}), but
now we combine two consecutive time steps.
We flip one node and look how the Hamming distance grows in comparison
to the undisturbed system after two time steps. The condition that
information spreading is critical is in the cycle with period 2
\begin{equation}
\pi_1(x) \cdot \pi_1(y) = \frac{1}{4}  \label{PiProd}
\end{equation}
Combining Eq.~(\ref{PiProd}) with the time evolution of $x$ and $y$ we
obtain three equations
\begin{align*}
y &= 1 - [x^2(1-p_+)^2+(1-x)^2p_+^2+2x(1-x)p_+(1-p_+)] \\
x &= 1 - [y^2(1-p_+)^2+(1-y)^2p_+^2+2y(1-y)p_+(1-p_+)] \\
\frac{1}{4} &= (x+ p_+ -2 x p_+) (y + p_+ -2 y p_+) 
\end{align*}
This system can be solved numerically and we obtain a critical value
$p_{cn}=0.0657$. For $p_+$ below this value a perturbation at one node
will die out and all nodes will again blink with period 2. Above this
value, attractors must be longer than 2. 

\begin{figure}[thb!]
\includegraphics[width=0.475\textwidth]{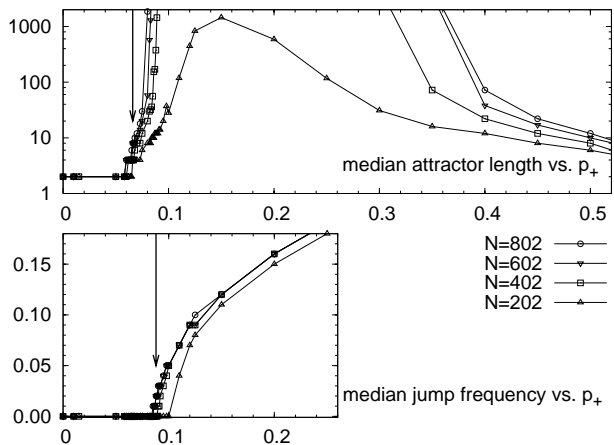}
\caption{Numerical verification for the transition~$p_{cn}$ and
$p_{cb}$, both marked by vertical arrows. The upper panel shows the
median attractor length, the lower panel the median jump frequency 
on each attractor candidate in dependence of $p_+$.
Each data point corresponds to 5000 sample networks of size $N$ with
fixed $p_+$ and two initial conditions per realization. The time
evolution is limited to 5000 computational steps for both the transient
and the attractor length. The median is therefore a more sensible
observable than the mean value: If a attractor candidate cannot be
verified because the time limit was already reached we can still
calculate the median.}
\label{TrbnSuccess} 
\end{figure}

We checked these analytical predictions by performing computer
simulations. In order to identify the transition at $p_{cn}$, we
measured the median attractor length. As shown in
Fig.~\ref{TrbnSuccess}, we find with increasing network size an
increasingly sharp transition. Below the transition at $p_{cn}$, the
proportion of attractors of length 2 converges to some nonzero value
with increasing system size, indicating that cycles of length 2 are
stable. Above the transition, the median attractor length increases
more and more rapidly with increasing system size, indicating a
diverging median. Another finding is that attractors become again
shorter as the critical point $p_+=1/2$ is approached.

We also evaluated the frequency of phase jumps in $b(t)$ on the
attractors.  Each phase jump is a deviation from an oscillation with
period 2.  The result is shown in the lower part of
Fig.~\ref{TrbnSuccess}. With increasing system size, there is an
increasingly sharp transition at $p_{cb}$ between zero phase jumps and
a finite proportion of phase jumps. 

We summarize the different types of dynamical behavior in the following
diagram, Fig.~\ref{PhaseDiagram}. 
\begin{figure}[htb]
\begin{tabular}{p{1.75cm}p{1.75cm}p{1.75cm}p{1.75cm}}
\hline 
\multicolumn{3}{|l}{nonfrozen} & \multicolumn{1}{|l|}{frozen} \\ \hline 
\multicolumn{2}{|l}{$b$ has period 2} &
 \multicolumn{2}{|l|}{$b$ is constant in time} \\ \hline
\multicolumn{1}{|p{2cm}}{all nodes have period 2} & 
 \multicolumn{3}{|l|}{nodes oscillate differently or not at all}\\ \hline
$p_+$ & $p_{cn}$ & $p_{cb}$ & $0.5$ \\
\end{tabular}
\caption{Overview of the dynamic behavior of the model with $K=2$ in
dependence of the parameter~$p_+$.}
\label{PhaseDiagram}
\end{figure}

To conclude, we have shown that the simplest Threshold Random Boolean
Network shows three different types of phase transitions and not just
the generally expected transition between a frozen and a chaotic
phase. For parameter values $p_+< p_{cn}$, all nodes oscillate stably
with period two. For $p_{cn}<p_+<p_{cb}$, the fraction of on-nodes
oscillate with period two, but attractors are longer.  For
$p_{cb}<p_+<1/2$, the dynamical behavior is chaotic in the sense
defined by Kauffman. For $p_+>1/2$, the network is in the frozen
phase.

The lesson to be learned from this study is that the dynamical
behavior of Boolean networks can be much richer than expected from
simple considerations. There is indeed no reason why some model should
not also show global oscillations with higher periods or period
doubling cascades in the temporal behavior of $b(t)$. Real genetic
networks can be expected to have an even richer dynamical behavior. If
the simple classification into ``frozen'', ``critical'' and
``chaotic'' networks fails already in the random model presented in
this paper, it will be even less suitable for real genetic networks,
which have attractors with very specific properties related to the
function of the network. A more sophisticated way of describing and
classifying the dynamical behavior of Boolean networks is therefore
required. 

\begin{acknowledgements}
\noindent This work was supported by the Deutsche Forschungsgemeinschaft
(DFG) under Contract No. Dr200/4-1.
\end{acknowledgements}

\bibliography{fgTrbnBib}
\end{document}